\documentclass[sigconf]{acmart}
\copyrightyear{2020}
\acmYear{2020}
\setcopyright{acmcopyright}
\acmConference[SIGIR '20] {43rd International ACM SIGIR Conference on Research and Development in Information Retrieval}{July 25--30, 2020}{Virtual Event, China}
\acmBooktitle{43rd International ACM SIGIR Conference on Research and Development in Information Retrieval (SIGIR '20), July 25--30, 2020, Virtual Event, China}
\acmPrice{15.00}
\acmDOI{10.1145/3397271.3401310}
\acmISBN{978-1-4503-8016-4/20/07}

\usepackage{subfigure}

\usepackage{appendix}

\AtBeginDocument{%
  \providecommand\BibTeX{{%
    \normalfont B\kern-0.5em{\scshape i\kern-0.25em b}\kern-0.8em\TeX}}}

\begin{document}
\fancyhead{}


\title{Deep Interest with Hierarchical Attention Network for Click-Through Rate Prediction}

\author{Weinan Xu, Hengxu He}
\authornote{This author deserves first author equal contribution.}
\author{Minshi Tan, Yunming Li, Jun Lang, Dongbai Guo}

\affiliation{Alibaba Group \\ \{stella.xu, minshi.tan, yunming.li, bill.lang, dongbai.guo\}@lazada.com, hengxu.hhx@alibaba-inc.com}

\begin{abstract}

Deep Interest Network (DIN) is a state-of-the-art model which uses attention mechanism to capture user interests from historical behaviors. User interests intuitively follow a hierarchical pattern such that users generally show interests from a higher-level then to a lower-level abstraction. Modelling such interest hierarchy in an attention network can fundamentally improve the representation of user behaviors. We therefore propose an improvement over DIN to model arbitrary interest hierarchy: Deep Interest with Hierarchical Attention Network (DHAN). In this model, a multi-dimensional hierarchical structure is introduced on the first attention layer which attends to individual item, and the subsequent attention layers in the same dimension attend to higher-level hierarchy built on top of the lower corresponding layers. To enable modelling of multiple dimensional hierarchy, an expanding mechanism is introduced to capture one to many hierarchies. This design enables DHAN to attend different importance to different hierarchical abstractions thus can fully capture a user’s interests at different dimensions (e.g. category, price or brand). To validate our model, a simplified DHAN is applied to Click-Through Rate (CTR) prediction and our experimental results on three public datasets with two levels of one-dimensional hierarchy only by category. It shows DHAN’s superiority with significant AUC uplift from 12$\%$ to 21$\%$ over DIN. DHAN is also compared with another state-of-the-art model Deep Interest Evolution Network (DIEN), which models temporal interest. The simplified DHAN also gets slight AUC uplift from 1.0$\%$ to 1.7$\%$ over DIEN. A potential future work can be combination of DHAN and DIEN to model both temporal and hierarchical interests.

\end{abstract}

\ccsdesc[500]{Information systems~Users and interactive retrieval}
\ccsdesc[500]{Information systems~Personalization}

\keywords{Click-Through Rate Prediction; Hierarchical Pattern; Hierarchical Attention Network; Recommendation}

\maketitle

\section{Introduction}
Click-Through Rate (CTR) prediction has become a core task in e-commerce since it is directly related to revenues. Furthermore, CTR also influences user satisfaction. Thus, CTR prediction modeling has drawn more attention from both academia and industry. 

Many CTR prediction models such as Wide and Deep (WDL) \cite{cheng2016wide} and Product Neural Network (PNN) \cite{qu2016product} have used deep learning methods to extract item-level features and feature interactions. However, these models do not consider to capture user interests through historical behaviors. As the success of attention-based models , several state-of-the-art CTR prediction models \cite{zhou2018deep,zhou2019deep} have been proposed to capture user interests by extracting behavioral feature based on user historical behaviors. Deep Interest Network (DIN) \cite{zhou2018deep}, the first work to use attention-based model in recommendation systems, indicates the diversity of user interests, and uses attention mechanisms to activate historical behaviors with respect to the target item.

However, user interests intuitively follow a hierarchical pattern such that users generally show interests from a higher-level attributes (e.g. category, price or brand) and progressively to a lower-level attributes or items while DIN and other state-of-the-art models overlook it.

To tackle the above issue, 
a novel structure Deep Interest with Hierarchical Attention Network (DHAN) is proposed to model arbitrary interest hierarchy. In DHAN, a multi-dimension with arbitrary depth hierarchical structure is constructed such that horizontally, the structure captures interests in various dimensions (e.g. category, price or brand); vertically, for each dimension, the structure captures hierarchical interests. Specifically, the multi-dimensional structure is introduced on the first attention layer attending to individual item by different dimensions, and the hierarchical structure is introduced by subsequent attention layers in the same dimension to attend higher-level hierarchy built on top of the lower corresponding attention layers. In this way, DHAN can attent to attributes of different levels of various dimensions with different importance when constructing the overall interest representation.

Deep Interest Evolution Network (DIEN) \cite{zhou2019deep} is another state-of-the-art model which considers dynamic changes of interests in time horizons. DHAN and DIEN focus on two different pathways to model interests. 
We also conduct experiments to compare the effects of two different pathways towards CTR prediction. 


The main contributions of this paper are summarized as follows:
\begin{itemize}
\setlength{\itemsep}{4pt}
\setlength{\parskip}{-2pt}
\item A novel structure DHAN is proposed to model arbitrary interest hierarchy with its scalability both on dimensions and hierarchical levels. The simplified structure with two levels and one dimension has achieved significant uplift on AUC compared with DIN over three public datasets. 
\item Compared with DIEN, the simplified DHAN has achieved slight uplift on AUC. The comparisons of modeling interest hierarchy over evolution indicate a potential future work to model both temporal and hierarchical interests.

\end{itemize}

\section{Deep Interest with Hierarchical Attention Network}
In this section, we introduce Deep Interest with Hierarchical Attention Network (DHAN) in detail. Firstly, we introduce the structure of DHAN. Secondly, structures of simplified DHAN referring to DIN and DIEN are presented. 

\subsection{The Structure of DHAN}
\begin{figure}[t]
  \centering
  \includegraphics[width=3.5in,height=3in]{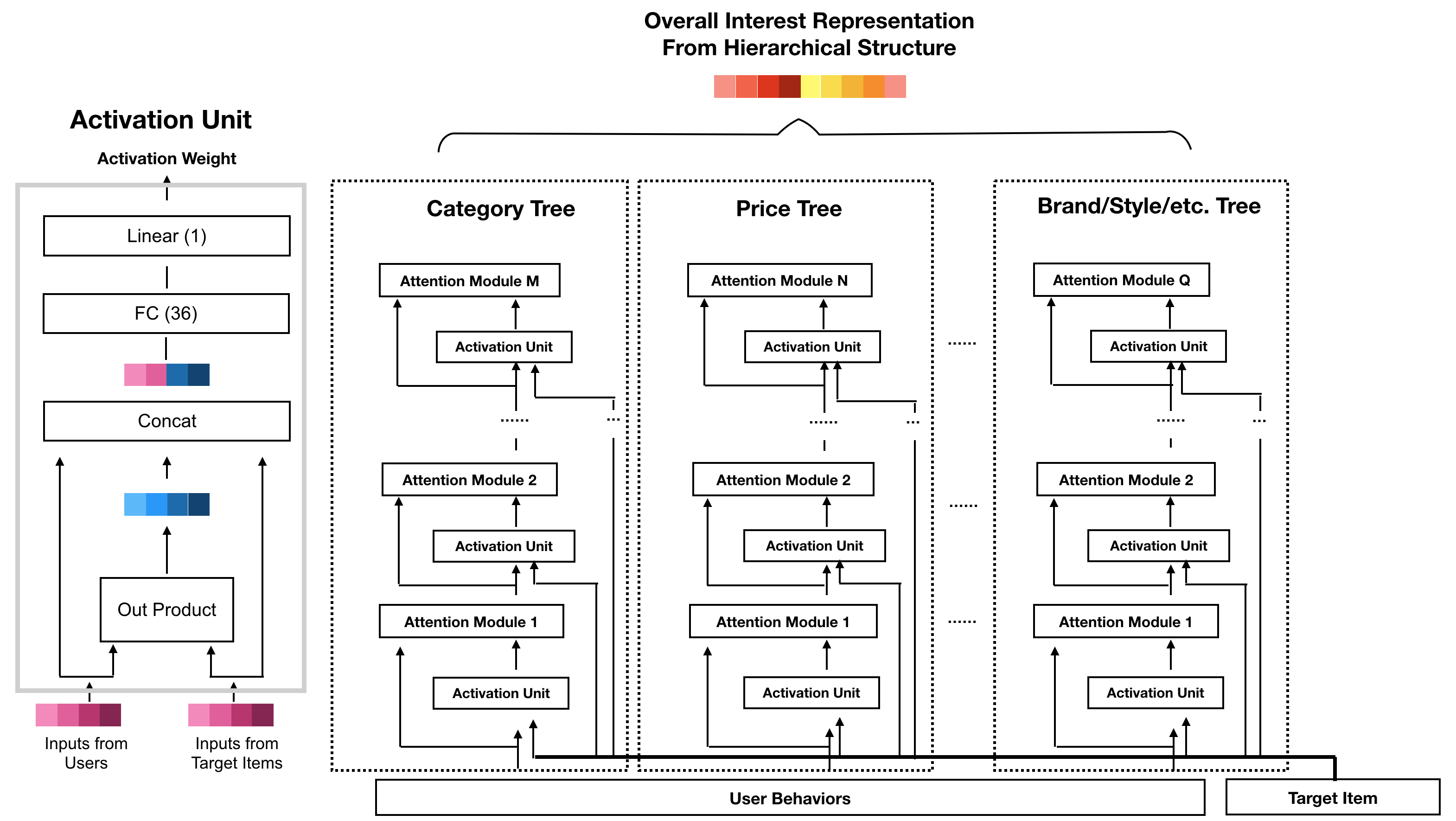}
  \setlength{\abovecaptionskip}{-0.1cm}
  \setlength{\belowcaptionskip}{-0.7cm}
  \caption{The main structure of DHAN. A multi-dimensional structure is introduced on the Attention Module 1. Each attention module is achieved based on the activation unit. For each tree, higher attention modules are built on top of the lower corresponding attention modules.}
  \label{fig:HAN_all}
\end{figure}

In order to model interest hierarchy with multiple dimensions and arbitrary depth, Deep Interest with Hierarchical Attention Network (DHAN) is proposed. As shown in Fig. \ref{fig:HAN_all}, multiple dimensions can be involved such as category, price, brand and style. The expanding mechanism is introduced above user behaviors and the target item in order to capture interests from one to many dimensions. For each dimension, the interest hierarchical structure is modeled like a tree from the lower-level attributes or items onto the higher-level attributes. Because of the hierarchical structure, the higher-level features are extracted based on the corresponding lower-level features by a series attention modules, enabling DHAN to attend different importance to different attribute levels. For each attention module, the activation unit shown in Fig. \ref{fig:HAN_all} is used for activating user interested items with respect to the given target item. On the top level, the overall interest representation incorporates user interests extracted from all levels and dimensions.

\subsection{The Structure of Simplified DHAN}
Considering the deployment of DHAN in recommendation scenarios, the simplified DHAN is used with one dimension and two attention levels.

In our experiments on public datasets, we use two categories of feature: \emph{User Behaviors} and \emph{Target Item}. The fields of \emph{User Behaviors} are the list of user \emph{reviewed item ids}, and the embedding of user behaviors is represented as $E=\{e_1,e_2,\cdots,e_t,\cdots,e_T\}$, where $e_t \in {R^d}$, $T$ is the number of user's historical behaviors. \emph{Target Item} is the item to recommend. The field of \emph{Target Item} is its \emph{item id}, and the embedding of target item is $e_a$ where  $e_a \in {R^d}$. The loss function is the negative log-likelihood function defined as: 
\begin{equation}
\setlength{\abovedisplayskip}{3pt}
\setlength{\belowdisplayskip}{3pt}
L =  - \frac{1}{N}\sum\limits_{(x,y) \in S}^{} {(y\log p(x) + (1 - y)\log (1 - p(x)))}, 
\end{equation}
where $S$ is the training set with data size $N$, and $x$ is the last layer of feature extraction before passing into fully-connected layers. $y \in \{0,1\}$ is the label of data, and $p(x)$ is the output after softmax, which is the probability of reviewing on sample $x$.

\begin{figure}[t]
  \centering
  \includegraphics[width=3.5in,height=2.5in]{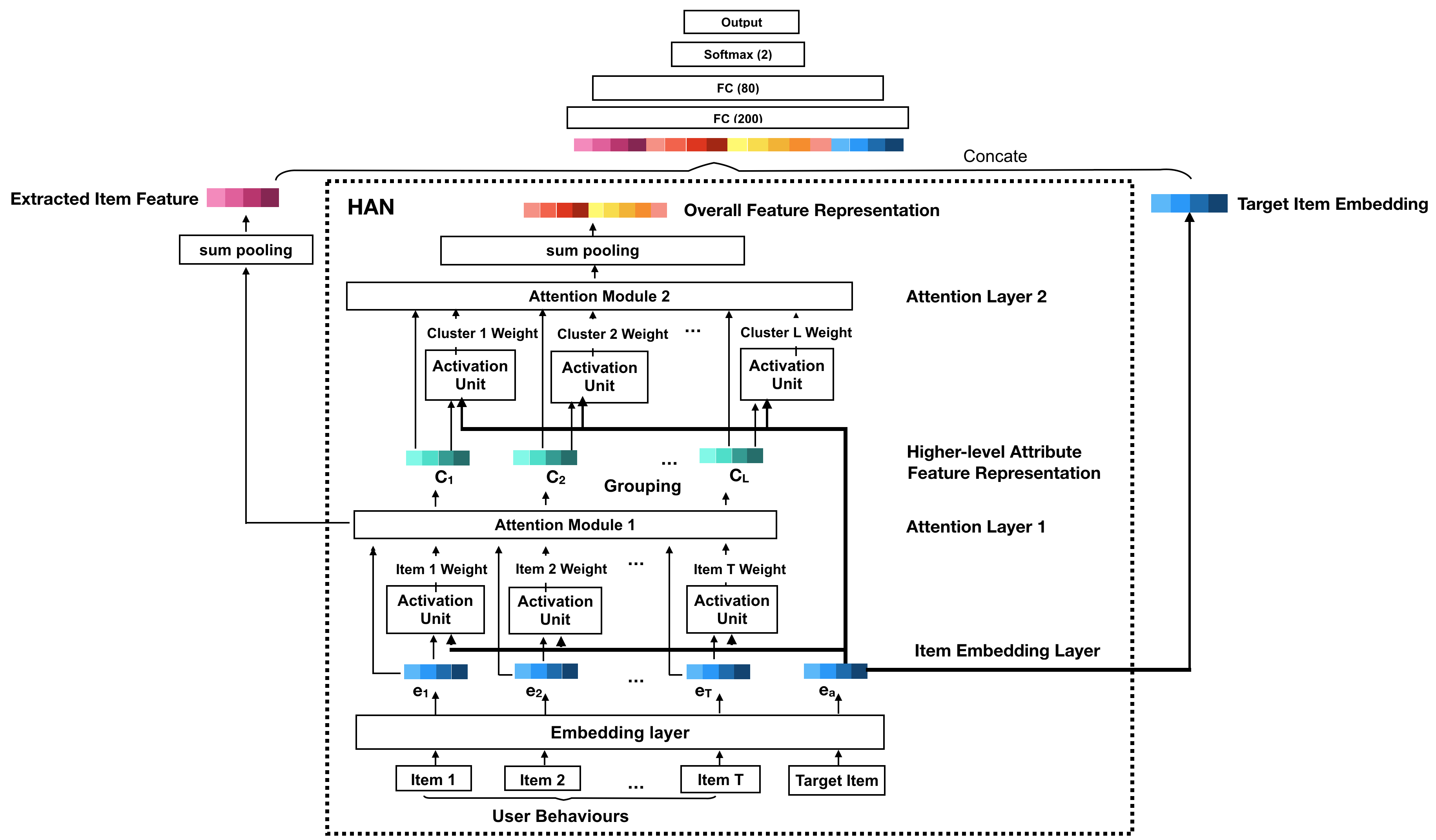}
  \setlength{\abovecaptionskip}{-0.2cm}
  \setlength{\belowcaptionskip}{-0.5cm} 
  \caption{The structure of simplified DHAN, and the main structure HAN is marked.}
  \label{fig:DHAN}
\end{figure}

The main part of the simplified DHAN is the Hierarchical Attention Network (HAN) marked in Fig. \ref{fig:DHAN}. HAN is to model interest hierarchy with two levels which are item level and attribute level. The item-level representation is the item embedding $E$. According to DHAN, the representations of attribute levels and overall interests are generated based on two attention modules. In Fig. \ref{fig:DHAN}, the first attention module is applied such that item weights $W=\{w_1,w_2,\cdots,w_t,\cdots,w_T\}$ are calculated by attention mechanism through activation units. For HAN, item embedding and item weights are grouped based on their attributes to generate each attribute-level feature representation. Let the clustered item embedding be $I=\{I^1, I^2,\cdots I^l, \cdots, I^L\}$ with $L$ groups obtaining from $E$, and each $I^l$ is $I^l=\{i_1^l,\cdots,i_X^l\}$ where $X$ is the length of user behaviors under group $l$. The corresponding weight of $I^l$ is $w^l=\{w_1^l,\cdots,w_X^l\}$ obtaining from $W$. To capture the user interests under each higher-level attribute $l$, each higher-level attribute feature is represented as $c^l$, which is a normalized weighted sum:
\begin{equation}
\setlength{\abovedisplayskip}{3pt}
\setlength{\belowdisplayskip}{3pt}
{c^l} = {{(w_1^li_1^l +  \cdots  + w_X^li_X^l)}}{{/\sum\limits_{k = 1}^X {w_k^l} }}.
\end{equation}

\begin{figure}[t]
  \centering
  \includegraphics[width=3.5in,height=2.5in]{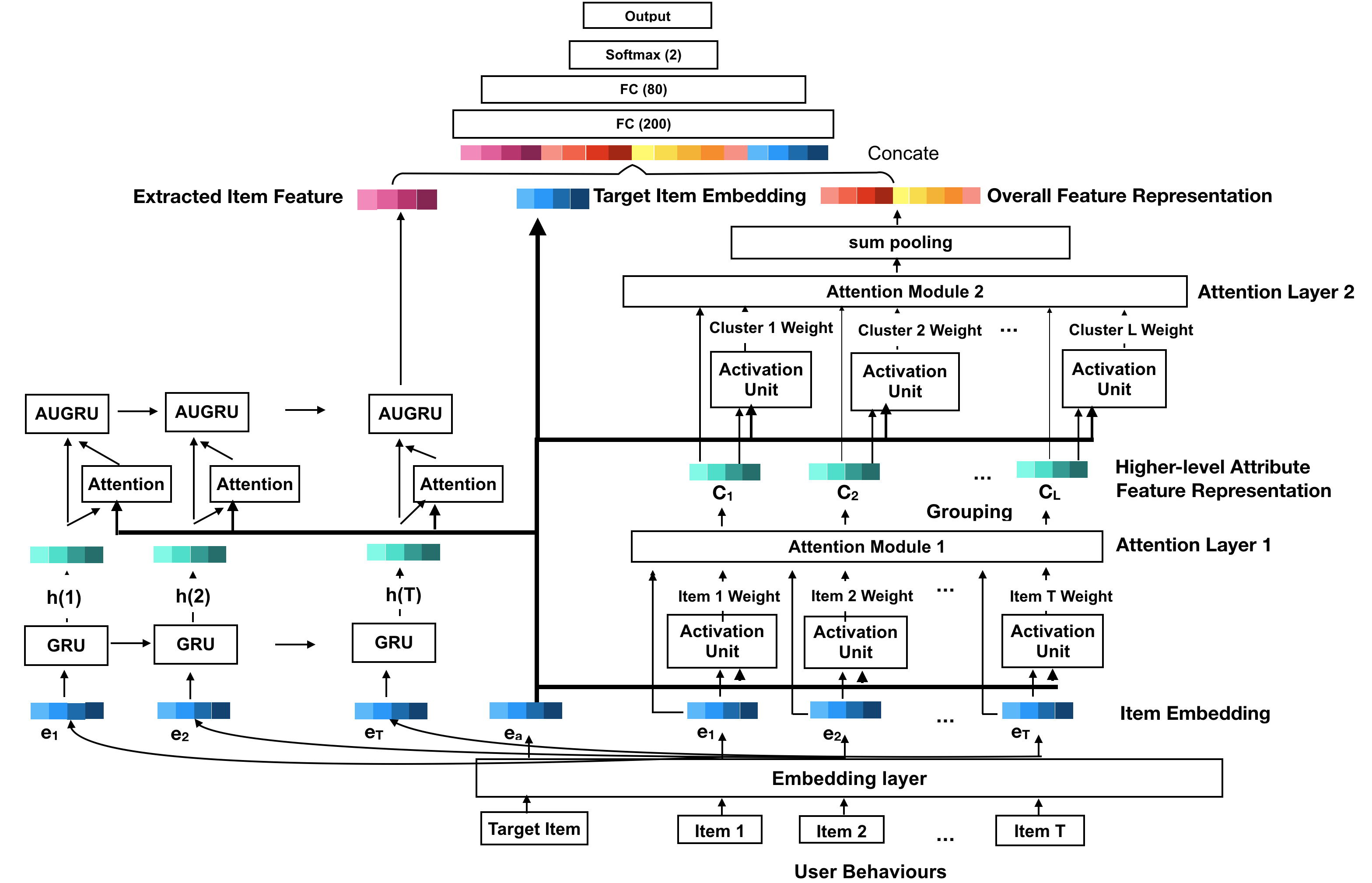}
  \setlength{\abovecaptionskip}{-0.3cm}
  \setlength{\belowcaptionskip}{-0.5cm}
  \caption{The structure of simplified DHAN which includes the extracted item feature from DIEN.}
  \label{fig:DHEAN}
  
\end{figure}

The second attention module is applied on top of the attribute level to extract overall interest representation under one dimension. The cluster weights of attributes are obtained by the second attention layer through the activation unit as $W_c=\{w_{c^1},w_{c^2},\cdots,w_{c^l},\cdots\\,w_{c^L}\}$.
The overall interest representation $x$ is obtained by the attention mechanism as 
$x = \sum\limits_{l = 1}^L {{w_{c^l}}{c^l}}$ .

After passing through the two attention modules, the output of HAN is the overall feature $x$. Compared with DIN \cite{zhou2018deep} where $x$ is extracted based on only one item-related attention layer, $x$ from HAN is extracted based on a hierarchical pattern by two attention layers. Thus, $x$ from HAN can contain more information than that from DIN. However, if the number of higher-level attributes is small, $x$ from HAN may have the over-centralized issue so that the capturing of interests is over concentrated to some higher-level attributes at early stage, causing the loss function trapped in local optima. To alleviate this issue, and to enhance the capability of feature expression before feeding into MLPs, $x$ is concatenated with target item embedding $e_a$ and item-level feature representation $i_x$. Accordingly, for the comparison of DHAN with DIEN, the item-level feature representation $i_x$ uses the same $i_x$ as DIEN with interest extraction layers by GRUs shown in Fig. \ref{fig:DHEAN}. Thus, the overall interest representation is $X=[x,i_x,e_a]$, which is then fed into the following MLP for final CTR prediction. 

\section{Experiments}

In this section, we present our experiments in detail, including experimental setup, experiment results and analysis. We compare our model simplified DHAN with the state-of-the-art models in public datasets. For simplicity, we use DHAN to notate simplified DHAN in this section. DIN is set as our baseline model in order to study the effect of including the extracted overall feature from the hierarchical structure compared with considering only the item-level extracted feature. Additionally, the model comparison between DHAN and DIEN is also conducted with the same public dataset but a different sample generation method because DIEN considers user interests evolution in time horizons. Besides, Wide and Deep (WDL) model and PNN model are also involved for comparisons because they are widely used in Embedding\&MLP. Each experiment is repeated 5 times, and mean and standard deviation are reported.

RelaImpr\cite{RelaImpr} metric is used to measure the relative improvement over models, which is defined as: 
\begin{equation}
\setlength{\abovedisplayskip}{3pt}
\setlength{\belowdisplayskip}{3pt}
{RelaImpr} = \left(\frac{AUC(measured~model)-0.5}{AUC(base~model)-0.5}-1 \right) \times 100\%.
\end{equation}

\subsection{Comparison with DIN}

\textbf{Experimental Setup}
\

Amazon Dataset is used as our benchmark, which contains product review information and metadata from Amazon. Our task is to predict (k+1)-th reviewed item by using the first k reviewed items. For each user, we construct positive sample by taking user's historical reviews, and generate a negative sample from items not reviewed by this user. Training and testing data sampled from three different datasets: Six-Category, Kindle Shop, and Electronics are used for the performance evaluation of DHAN. The Six-Category contains six subsets: Video Games, Movies and TV, Industrial and Scientific, Electronics, CDs and Vinyl, Apps for Android, and we pick their category names as our category tags and only take 10\% of reviews. For the other two sampled datasets, we only keep categories that contain more than 300 items to make the datasets more concentrated, and evaluate the model performance on digital and  physical goods separately.

\begin{table}
  \setlength{\abovecaptionskip}{0.cm}
  \caption{The statistics of datasets with DIN-generation.}
  \label{tab:freq}
  \footnotesize
  \begin{tabular}{lp{0.8cm}p{0.8cm}p{1.2cm}p{0.8cm}}
    \toprule
    Dataset & User & Goods & Categories & Samples \\
    \midrule
    Six-Category & 259541 & 138454 & 6 & 463624\\
    Kindle Shop & 65599 & 40241 & 48 & 1221635\\
    Electronics & 188757 & 33000 & 51 & 1286668\\
  \bottomrule
\end{tabular}
\vspace{-0.5cm}
\end{table} 

\noindent
\textbf{Experiment Results and Analysis}
\

Table~\ref{tab:DIN} describes the results of DHAN compared with DIN and Embedding\&MLP on public datasets. Here, DIN is treated as the state-of-the-art baseline. Although DIN beats basic Embedding\&MLP (WDL and PNN) with its capability of finding user interested items by attention mechanisms, DHAN beats DIN since firstly, DHAN captures the overall feature representation of user interests with the hierarchical pattern, and secondly, DHAN combines the overall features with item-level features to achieve effective interest expression. 
The model with hierarchical attention network helps DHAN to beat DIN with remarkable AUC improvement, i.e. 12.58\% RelaImpr on Amazon Six-Category Dataset, 14.79\% on Kindle Shop Dataset and 21.03\% on Electronics Dataset.

\begin{table}
\setlength{\abovecaptionskip}{0.cm}
  \Huge
  \caption{Results on Amazon datasets, with DIN as baseline.}
  \label{tab:DIN}
  \resizebox{\linewidth}{!}{
  
  \begin{tabular}{ccccccc}
    \toprule
    Model&\multicolumn{2}{c}{Amazon (Six-Category)}&\multicolumn{2}{c}{Amazon (Kindle Shop)}&\multicolumn{2}{c}{Amazon (Electronics)}\\
         & $mean\pm std$&RelaImpr& $mean\pm std$&RelaImpr& $mean\pm std$&RelaImpr\\
    \midrule
     WDL & $0.6949\pm 0.00000$ & $-11.49\%$ & $0.8928\pm 0.00032$ & $-0.10\%$ & $0.8420\pm 0.00115$ & $-5.14\%$\\
     PNN & $0.6999\pm 0.00046$ & $-9.23\%$ & $0.8363\pm 0.00069$ & $-14.49\%$ & $0.8498\pm 0.00066$ & $-2.96\%$\\
     DIN & $0.7202\pm 0.00007$ & $0.00\%$ & $0.8932\pm 0.00090$ & $0.00\%$ & $0.8605\pm 0.00054$ & $0.00\%$ \\
     DHAN & $0.7479\pm 0.00010$ & $12.58\%$ & $0.9514\pm 0.00291$ & $14.79\%$ & $0.9363\pm 0.00111$ & $21.03\%$\\
  \bottomrule
\end{tabular}
}
\vspace{-0.3cm}
\end{table}

\begin{figure}[h]

  \centering
  \includegraphics[width=2.5in]{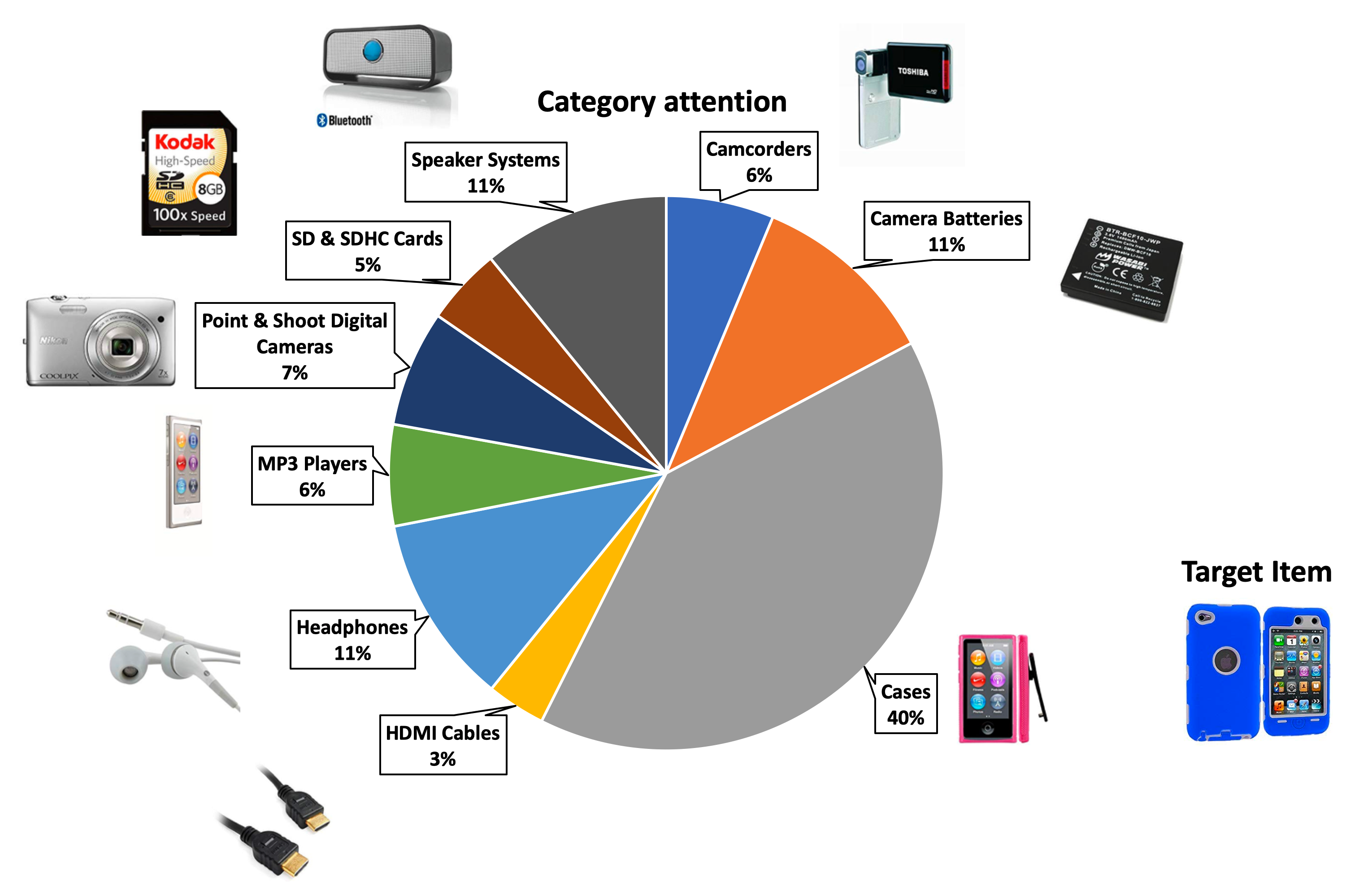}
   \setlength{\abovecaptionskip}{-0.cm}  
   \setlength{\belowcaptionskip}{-0.7cm}  
  \caption{Visualization of higher-level activation in DHAN. This user has the highest interests on Cases category, which is the most relevant category to the target item.}
  \label{fig_cat_attention}
\end{figure}

\begin{figure}[htbp]
\centering
\subfigure[Training\_Loss]{
\begin{minipage}[htbp]{0.4\linewidth} 
\centering
\includegraphics[width=1.2in]{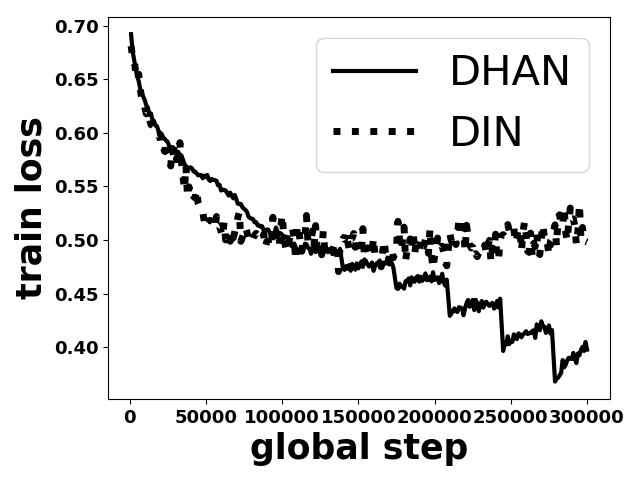} 
\end{minipage}
}
\subfigure[AUC]{
\begin{minipage}[htbp]{0.4\linewidth}
\centering
\includegraphics[width=1.2in]{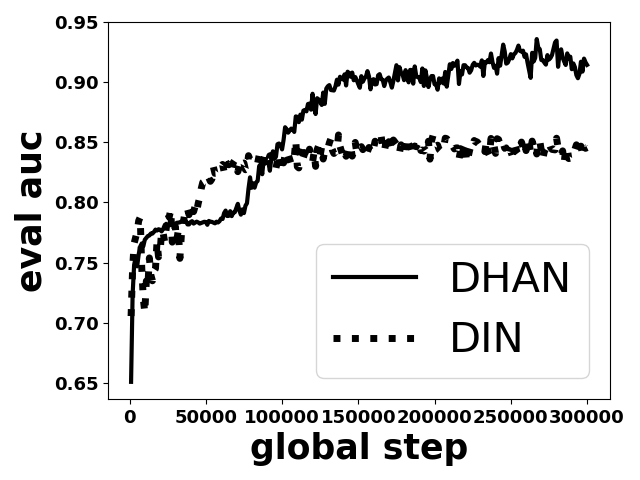}
\end{minipage}
}%

\centering
\setlength{\abovecaptionskip}{-0.1cm}
\setlength{\belowcaptionskip}{-0.7cm}
\caption{Training loss and AUC of DHAN and DIN on the Amazon Electronics Dataset.}
\label{loss and auc}
\end{figure}

The effect of the higher-level activation is visualized in Fig. \ref{fig_cat_attention}, which illustrates interest intensity of higher-level attributes (category in the case) with respect to the target item. Obviously, the higher-level activation attends to high relevance to the target item with the attention weight as 40$\%$ on handphone cases, which support the effect of the hierarchical structure in DHAN. Fig. \ref{loss and auc} shows the comparison of DHAN and DIN. After 100000 global steps, DHAN shows its significant uplift on both AUC and training loss.

\subsection{Comparison with DIEN}
\textbf{Experimental Setup}
\

Similar to 3.1, we still use Electronics, Kindle Shop, and Six-Category datasets from Amazon to construct our training and testing samples. To keep in accordance with the DIEN sample generation method which dedicates to extracting sequential features, each users' sequential review history is taken as one record if the user has at least 5 reviewed items. Due to the different criteria of selecting samples from each dataset, the statistics in Table \ref{tab:freq_dien} is different from those in Table \ref{tab:DIN}. A negative sample is generated by selecting items not reviewed by each user. Since the sequential review history is taken based on each user in DIEN, when sampling in Six-Category dataset, we take 10\% users' review history, instead of 10\% of entire review history directly. 

\noindent
\textbf{Experiment Results and Analysis}
\

Table~\ref{tab:DIEN} shows the results on public datasets where DIEN is the state-of-the-art baseline. Among all the attention-based models, DIEN performs better than DIN since DIEN takes interests evolution in time horizon into account. However, DHAN finally beats DIEN by 1.09\% on Amazon Six-Category, 1.75\% on Kindle Shop and 1.50\% on Electronics. The improvement of DHAN is slight because DIEN proposes a interest extractor layer to capture temporal interest from historical behaviors, but DHAN does not capture this feature. In addition, training and testing sample generation method to adapt to the sequential behaviors in DIEN reduces the AUC gap between DHAN and DIN. This will be addressed in future work.

\begin{table}
\setlength{\abovecaptionskip}{0.cm}
  \caption{The statistics of datasets with DIEN-generation.}
  \label{tab:freq_dien}
  \footnotesize
  \begin{tabular}{lp{0.8cm}p{0.8cm}p{1.2cm}p{0.8cm}}
    \toprule
    Dataset & User & Goods & Categories & Samples \\
    \midrule
    Six-Category & 49673 & 139531 & 6 & 99346\\
    Kindle Shop & 46776 & 40221 & 48 & 93552\\
    Electronics & 73958 & 33175 & 51 & 147916\\
  \bottomrule
\end{tabular}
\end{table}

\begin{table}
\setlength{\abovecaptionskip}{0.cm}
  \Huge
  \caption{Results on Amazon datasets, with DIEN as baseline.}
  \label{tab:DIEN}
  \resizebox{\linewidth}{!}{
  \begin{tabular}{ccccccc}
    \toprule
    Model&\multicolumn{2}{c}{Amazon (Six-Category)}&\multicolumn{2}{c}{Amazon (Kindle Shop)}&\multicolumn{2}{c}{Amazon (Electronics)}\\
         & $mean\pm std$&RelaImpr&$mean\pm std$&RelaImpr&$mean\pm std$&RelaImpr\\
    \midrule
    WDL & $0.9011\pm 0.00357$& $-3.41\%$ &$0.7936\pm 0.00292$&$-18.49\%$ &$0.7023\pm 0.00034$&$-9.34\%$\\
    PNN & $0.9127 \pm 0.00030$ & $-0.63\%$ & $0.8215\pm 0.00068$&$-10.74\%$&$0.7123\pm 0.00154$&$-4.82\%$\\
    DIN & $0.9075\pm 0.00000$ & $-1.88\%$ & $0.8321\pm 0.00011$&$-7.81\%$&$0.7138\pm 0.00000$&$-4.17\%$\\
    DIEN & $0.9153\pm 0.00000$ & $0.00\%$ & $0.8602\pm 0.00000$ & $0.00\%$ &$0.7231\pm 0.00000$& $0.00\%$\\
    DHAN & $0.9198\pm 0.00167$ & $1.09\%$ &$0.8665\pm 0.00079$ &$1.75\%$&$0.7264\pm 0.00148$&$1.50\%$\\
  \bottomrule
\end{tabular}}
\end{table}

\section{Conclusion}
In this paper, a novel model Deep Interest with Hierarchical Attention Network (DHAN) is proposed to model user interest hierarchy for multiple dimensions and arbitrary depth. An expanding mechanism in the item level is introduced to capture one to many dimensions. For each dimension, the higher-level features are extracted based on the corresponding lower-level features by a series of attention modules. Considering the deployment in recommendation scenarios, a simplified DHAN has been applied to three public datasets, and has achieved significant uplift of AUC around 12$\%$ to 21$\%$ over DIN and slight uplift of AUC around 1.0$\%$ to 1.7$\%$ over DIEN. The outperformance of DHAN supports the effectiveness of modeling interest hierarchy leads to more accurate CTR prediction. The differences in uplifts are because DIEN models interest evolution. Thus, a potential future work is to model user interests considering both interest hierarchy and evolution. Futhermore, as more dimensions are involved in item features, another future work is to replace the predefined dimensions and automatically capture user interests towards those dimensions.

\bibliographystyle{ACM-Reference-Format}
\bibliography{DHAN_SIGIR_final}

\appendix
\section{Open source codes}
Our open-source codes with parameters are available on GitHub: \\https://github.com/stellaxu/DHAN.

\end{document}